\documentclass{IEEEtran}

\usepackage{cite}
\usepackage{amsmath,amssymb,amsfonts}
\usepackage{algorithmic}
\usepackage{graphicx}
\usepackage{textcomp}
\usepackage{xcolor}
\usepackage{mathtools}
\usepackage{amssymb}
\usepackage{braket}
\usepackage{nicefrac}
\usepackage{xfrac}
\usepackage{caption}
\usepackage{subcaption}
\usepackage{multicol}
\usepackage{float}
\usepackage{flushend}
\usepackage{xcolor, authblk, microtype}
\DeclareMathOperator*{\argmax}{argmax}
\def\BibTeX{{\rm B\kern-.05em{\sc i\kern-.025em b}\kern-.08em
    T\kern-.1667em\lower.7ex\hbox{E}\kern-.125emX}}

\title{On Sampling and Inference using Quantum Algorithms\\
}

\makeatother
\author[1]{S Ashutosh}
\author[2]{Deepankar Sarmah}
\author[3]{Sayantan Pramanik}
\author[4]{M Girish Chandra}

\affil[1,2]{Indian Institute of Science Education and Research, Kolkata}

\affil[3,4]{TCS Research and Innovation, India}
\makeatletter
\renewcommand\AB@affilsepx{, \protect\Affilfont}
\affil[1,2]{\textit {\{sa17ms105, ds15ms006\}@iiserkol.ac}}\affil[3,4]{\textit {\{sayantan.pramanik, m.gchandra\}@tcs.com}}

\begin{document}
\maketitle

\begin{abstract}
Quantum computers are projected to handle the Gibbs sampling and the
related inference on Markov networks effectively. Apart from noting the background
information useful for those starting the explorations in this important thread of
Quantum Machine Learning, we capture some results and observations obtained
through extensive simulations with two popular paradigms of sampling based on
Quantum Annealing and Quantum Approximate Optimization Algorithm.
\end{abstract}

\section{Introduction}\label{Introduction}
In this work, we delve into the problem of Gibbs sampling and using it for probabilistic inference on simple Markov networks. As far as sampling at different temperatures is concerned, the methods based on both Quantum Annealing (QA) and Quantum Approximate Optimization Algorithm (QAOA) are adopted. The paper borrows heavily from the existing resources, but an attempt is made to put the concepts and necessary details in a cohesive manner. In this sense, the novelty is almost negligible barring couple of example simulation results. The authors believe that the paper can provide useful background information to those planning to explore this important topic of Quantum Machine Learning.

\section{Equilibration and Thermalization in Quantum Systems}\label{eth}
An useful starting point is to capture few basic aspects related to the equilibration and thermalization in quantum systems. In doing so, we have extracted  extensively from the contents of \cite{wilming} and \cite{wittek}. If a system is in an equilibrium state that is well described by the Gibbs distribution it is referred to as thermalized; implied here is the fact that a system that thermalizes has to equilibrate. The Gibbs distribution is a probability distribution for the probabilities $p_{j}$ to find a system in a state with energy $E_{j}$. The probabilities $p_{j}$ are given by $p_{j} = \nicefrac{e^{-\beta E_{j}}}{Z}$, where the constant $Z = \sum_{j}e^{-\beta E_{j}}$ is called the partition function and ensures that the probabilities add up to unity and $\beta = \nicefrac{1}{k_{B}T}$ is the inverse temperature; $k_{B}$ is the Boltzmann constant. In quantum mechanics, the Gibbs distribution is represented by a \textit{mixed state} with probabilities $p_{j}$ to find the system in an eigenstate with energy $E_{j}$. This mixed state is usually called the Gibbs state.

A popular approach to arriving at the Gibb’s distribution in quantum systems is through what is called “Pure state quantum statistical mechanics” . The basic idea is that the universe is ideally in a pure state and by looking only at a \textit{small part} of the universe one sees a mixed state of that part, which is very close to the Gibbs state. It is also well known that the Gibb’s state and hence the equilibrium state has to be the one that maximizes its (small part’s) entropy while keeping the expectation values of all conserved quantities fixed. Little more elaboration on the entropy is in order for the quantum case, where the entropy plays a very different role: In classical statistical mechanics the entropy is not a property of the state of the system but a measure of our lack of knowledge about the state of the system. In contrast to this, in the case of quantum mechanics the Gibbs state results as the reduction of a pure state of a larger quantum system. The entropy is not a measure of lack of knowledge about the state of the system that is caused by our inability to completely measure the state of the system, but results from \textit{entanglement}. Even if we had perfect knowledge about the pure state of the universe, the entropy of a subsystem would be non-zero if the universe is not in a product state of the subsystem and the rest of the universe.

As mentioned earlier, equilibration is a necessary condition for thermalization. In classical physics, or generally in what is usually called a dynamical system, an equilibrium state is a state that does not change over time. Due to the unitary evolution of a closed quantum mechanical system, however, a pure state of a closed system will never equilibrate in the sense that it does not change over time at all. The only exception to this is the situation when the system starts in an eigenstate of its Hamiltonian. Therefore, considering small part of a larger system appears to be a good rationale in quantum mechanical systems. It is well studied and shown that most subsystems of large enough systems in fact do equilibrate. The guiding idea is to think of a small system with limited degrees of freedom that is in contact with a large \textit{heat bath} which has very many degrees of freedom. In this context, it is also useful to note that generally only certain local observables with a support only on a small part of the system can be measured, but not the state itself. It therefore makes sense to say that a subsystem equilibrates if all observables with support only on this subsystem equilibrate; An observable $A$ equilibrates if its expectation value $\braket{A} = Tr(A \rho_{S}(t))$ is close to its time-average for most times $t$. 

In the previous paragraph, $\rho_{S}(t)$ is the state of the subsystem in terms of the density matrix. Recollect that a density matrix is a way of writing a quantum state, whether it is a pure or mixed state. Mixed states are \textit{classical} probability deistributions over pure states. Formally, a mixed state is written as $\sum_{i} p_{i} \ket{\psi_{i}} \bra{\psi_{i}}$, where $\sum_{i} p_{i} = 1$, $p_{i} \geq 0$. This reflects our classical ignorance over the underlying quantum states. Density matrix formalism is indispensable in the study of thermalization as in understanding noise models and decoherence. Again, starting with the pure state $\Psi (t)$ of the total system, consisting of the bath and the subsystem, the density matrix of the total system is the usual $\rho (t) = \ket{\Psi (t)} \bra{\Psi (t)}$. The state of the subsystem $\rho_{S}$ is given by \textit{tracing out the bath}, so $\rho_{S} = Tr_{B} (\rho)$, and correspondingly, $\rho_{B} = Tr_{S} (\rho)$. The dynamics of the subsystem is governed by the dynamics of the total system as $\rho_{S}(t) = Tr_{B} (\rho(t))$. It is useful to also note that if the subsystem and bath are described by Hilbert spaces $H_{S}$ and $H_{B}$ respectively, the total Hilbert space is given by $H_{S} \otimes H_{B}$. The Hamiltonian (energy of the system) governing the dynamics of the total system may be written as $H = H_{S} + H_{B} + H_{SB}$, where $H_{S}$ and $H_{B}$ onle act on the system and bath, respectively, and $H_{SB}$ contains the interactions.  Coming to the dynamics of the system $\rho_{S}(t)$, it is again useful to stress the fact that the subsystem does not evolve under unitary evolution governed by its Hamiltonian $H_{S}$.

Coming to thermalization, we speak of thermalization of a (sub)-system if it equilibrates \textit{and} can be well described by the Gibbs state 
\begin{equation}\label{eq:Gibbs}
\rho_{Gibbs} = \nicefrac{e^{-\beta H_{S}}}{Z}
\end{equation}
where, $H_{S}$ is the Hamiltonian describing the dynamics of the subsystem; note that the subsystem Hamiltonian enters into the description of the thermalized state. It is of fundamental interest and importance to understand when and why the Gibbs state gives a good description of a system's state. Fortunately, equilibrium is a common feature of sussystems of large quantum systems and further, in these cases when the formula is used to calculate expectation values of observables for systems at finite temperature, the results are often in very good agreement with experiments. For generic large systems we can therefore expect that small subsystems can be well described by the Gibbs state.

Before ending this section, it is worth bringing into the context- what process is responsible for the small subsystems to approach thermal equilibrium. A widely accepted mechanism is based on what is called as Eigenstate Thermalization Hypothesis (ETH). Any sub system that fulfills the ETH can locally be well described by the Gibbs state if it is in an eigenstate. In other words, according to ETH, thermalization occurs at the level of individual eigenstates of a given Hamiltonian (of the total system): Each eigenstate of the Hamiltonian implicitly contains a thermal state. More details can be found in \cite{wilming, eth}.

Unitary evolution is true for a closed system, that is, a quantum system perfectly isolated from the environment. This is not the case in the quantum computers we have today: these are open quantum systems that evolve differently due to to uncontrolled interactions with the environment.

\section{NISQ Open Systems}\label{NISQ}
Of late, we are witnessing spectacular developments in Quantum Information Processing with the availability of Noisy Intermediate-Scale  Quantum  devices  of  different  architectures \cite{qudo}. Two major groups of architectures are designed and experimentally tested: First one is the usage of quantum gates for unitary operations. A sequence of gates implements a given quantum circuit. The other group uses a more low-level approach. Adiabatic quantum computers (and their imperfect versions, the Quantum Annealers) leverage the Adiabatic Theorem \cite{babej} to preform transition from an lowest energy state of one system to the lowest energy state of another, under certain conditions. 

We would like a Quantum Processing Units (QPU) to be perfectly isolated from the environment, but in reality, the quantum computers we have today and for the next couple of years cannot achieve a high degree of isolation; they constantly interact with their environment in a largely uncontrolled fashion. This also means that their actual time evolution is not described by a unitary matrix as we would want it, but some other operator (the technical name for it is a completely positive trace-preserving map). Of course, as discussed earlier, even though QPU behaves as an open system, the QPU plus its associated environment (bath) evolves as a bigger closed system. If the environment is defined by a temperature $T$ , and if we let the system equilibrate, the QPU will become thermalized at temperature $T$. One can think of a cup of coffee: left alone, it will equilibrate with the environment, eventually reaching the temperature of the environment. This includes energy exchange. The energy of the thermal states of the QPU again follow Boltzmann distribution:
\begin{equation}
\rho_{S} = \frac{e^{-\beta H_{S}}}{Z}
\end{equation}
with $H_{S}$ being the Hamiltonian of the QPU with the partition function $Z = Tr(e^{\sfrac{-H_{S}}{T}})$ that $Tr(\rho_{S}) = 1$. If $H_{S}$ has a discrete basis of orthonormal eigenstates $\ket{n}$ with eigenvalues $E_{n}$, we can write $H_{S} = \sum_{n}E_{n} \ket{n} \bra{n}$ and 
\begin{equation}
\rho_{S} = \nicefrac{\sum_{n} e^{-\sfrac{E_{n}}{T}}\ket{n}\bra{n}}{Z}
\end{equation}
(since exponentiating a diagonal operator consists of exponentiating the elements of the diagonal). Hence, reiterating, the thermal density matrix is a mixed state where each eigenstate of $H_{S}$ where $E$ has a classical probability $P(E) = \frac{(e^-\sfrac{E}{T})}{Z}$, a Boltzmann distribution. We can see that the minimum energy eigenstate will have the highest probability. When $T \rightarrow 0$, the minimum energy eigenstate will have a probability close to $1$ and when $T \rightarrow \infty$, all the eigenstates tend to have equal probability.

The question that arises now is: how to approximate this thermalized state $\rho_{S}$ of the Hamiltonian using a quantum computing architecture? For pure ground states, there were two methods: QA and QAOA. It turns out that both of these methods can be adjusted to prepare thermalized density matrices.

\subsection{QA Gibb’s Sampler}\label{Gibbs}
Conventionally, as mentioned earlier, the QAs are designed in such a way that the lowest energy state of the initial system is easily attainable, while the lowest energy state of the target system encodes the output of the algorithm. Given the imperfect conditions of adiabatic quantum annealer, where it interacts with the environment, we also know that it is going to end up in the thermal state, following Boltzmann distribution. Thermalization being hard to simulate classically and this process could be exploited for calculations. Thus, every time when we face the problem of pulling out samples of a Boltzmann distribution, we can just plug in this idea of annealing to sample the thermal state. This is of special interest in certain areas of machine learning, since models such as Restricted Boltzmann Machines heavily rely on sampling in both training and usage phases.  In summary, the evolution of the annealer system can be used for our advantage if the viewpoint is changed from optimization to sampling. Research performed in the recent years indicates that samples obtained from a quantum annealer, such as D-Wave 2000Q, can produce samples from a \textit{low-temperature Boltzmann distribution}. One of the most important practical challenges with usage of the quantum annealers for sampling purposes is the fact that the device does not directly sample from the Boltzmann distribution with the temperature of the physical device or the dilution refrigerator that cools down the superconducting system (say, 0.0125K).  But, in fact sample from a distribution with a higher temperature referred to as the effective temperature; it is caused by noise and other factors that surround the quantum processing unit. So this has to be estimated and with this estimate we can rescale these energy values according to what the hardware is actually executing . See \cite{benedetti} for the importance of effective temperature. But, once this extra step is done, the sampling from the QA can be used as an algorithmic primitive. 

\subsection{Quantum Approximate Thermalization}\label{QAT}
There have been many machine learning applications where annealers has been seen to perform the task quite well, but an open ended question remained whether or not near-term circuit model quantum computers would be able to accomplish similar tasks. It is now established that we are not restricted to a quantum annealer and can actually do something very similar by using a gate model quantum computer. 

In regard to this Quantum Boltzmann Machines were proposed. These are a type of neural network model in which a network of spins representing bits of data are typically trained to associate a low-energy to the spin representations of a training data-set distribution. The approaches to train these Boltzmann machine rely on variationally optimizing the network’s energy function. This could be done by employing Quantum Approximate Optimization Algorithm (QAOA) \cite{qaoa} as a subroutine on shallow circuits in order to approximately sample from Gibbs states of Ising Hamiltonians (see Section \ref{ising}) and then using it for training. This is called quantum approximate thermalization \cite{verdon}. It is to be noted that even there were many results for preparing a thermal state on a gate-model quantum computer, most of them required a large-scale device. Only recently, a protocol for approximating thermalization was developed using shallow circuits.

To get the rationale for quantum approximate thermalization, we can start with the maximally entangled state $\ket{\phi^{+}} = \frac{1}{\sqrt{2}}(\ket{00}+\ket{11})$ and marginalize over one of the subsystems, it doesn't matter which subsystem; in other words, we take the partial trace in one of the subsystems. This leads to the maximally mixed state:
\begin{equation}
Tr[\ket{\phi^{+}}\bra{\phi^{+}}] = \frac{I}{2}
\end{equation}

The $\nicefrac{I}{2}$ density matrix obtained above by tracing out has only diagonal elements and they are equal: this is the equivalent way of writing a uniform distribution. We know that the uniform distribution has maximum entropy, and for this reason, a density matrix with this structure is called a maximally mixed state. In other words, we are perfectly ignorant of which elements of the canonical basis constitute the state. Further, the maximally mixed state is just the thermal state at infinite temperature. 

Taking this idea further, by having some larger system and tracing it out appropriately can one arrive at a thermal state? And the answer is affirmative. Let us make it more concrete. We will start by creating some easy system, for instance, $\sigma^{x}$ Hamiltonian acting on each site, i.e., $H_{0} = -\sum_{i}\sigma_{i}^{x}$. n the direction of creating a larger system, we consider ancilla qubits, equal in number to that of $H_{0}$. Restricting to one qubit for $H_{0}$ and another qubit for ancilla, we can consider the larger system with a state where the said qubits are entangled; recollect the importance of entanglement in connection with the thermalization in quantum systems from Section \ref{eth}. The state with the entangled qubits is: 
\begin{equation}
\ket{\psi} = c(e^{-\sfrac{1}{2T}}\ket{+}_{S}\ket{+}_{A} + e^{\sfrac{1}{2T}}\ket{-}_{S}\ket{-}_{A})
\end{equation}
where $\ket{\pm} = \frac{1}{\sqrt{2}}(\ket{0}\pm \ket{1})$, Which resembles the maximally entangled state but involving $\ket{+}$ and $\ket{-}$ states; the former is an equal superposition of $\ket{0}$ and $\ket{1}$, and the latter is the same thing with a minus sign.

The extra difference is that we have coefficients $e^{-\sfrac{1}{2T}}$ and $e^{\sfrac{1}{2T}}$ in front of the individual terms and the normalization factor $c$. Since we are interested in the actual system, we can trace out the ancilla system: 
\begin{equation}
Tr[\ket{\psi}\bra{\psi}] = \frac{e^{-\sfrac{H_{0}}{T}}}{Z} = \rho_{0}
\end{equation}
Resulting exactly the thermal state of the Hamiltonian $H_{0}$. This (thermal) state which is in equilibrium with the environment is easy to prepare. Now we can use, as a second step, the QAOA to go from $\rho_{0}$ to the thermal state of a system of interest $\rho_{1}$, for instance, the Ising model. Recollect that in usual QAOA, we approximate the ground state of a target Hamiltonian, starting from the ground state of a mixer Hamiltonian. We can actually perform the exact same optimization for approximating the thermal state of a target system, starting from the thermal state of some other system. Since QAOA approximates the adiabatic pathway, it should be a conservative change, so at the end of it, we would be close to the thermal state of the target Hamiltonian. Expressing slightly differently, since the QAOA approximates a unitary evolution, it is not going to change the thermal equilibrium. So we will always stay close to the thermal equilibrium. So, we will always stay close to the thermal equilibrium. Once we complete optimizing over this parametric circuit, we will be able to read out an approximate thermal state of the system of interest.

Elaborating further, if we know the thermal state is a pure state, we could apply QAOA to get to the thermal state of a target Hamiltonian. In order to negotiate the thermal state of of our sub system, the trick is to \textit{purify} $\rho_{0}$ \textit{on a larger Hilbert space}. If $H_{1}$ is our current Hilbert space, purifying a density matrix $\rho_{0}$ consists of finding a second Hilbert space $H_{2}$ such that there exists $\ket{\Psi} \in H_{1} \otimes H_{2}$ with $\rho_{0} = Tr_{H_{2}}(\ket{\psi}\bra{\psi})$, where $Tr_{H_{2}}$ is the partial trace taken over the second Hilbert space; in essence, we are marginalizing the probability distribution  (see also \cite{poulin}).

It can be shown that $\ket{\Psi} = \sqrt{2cosh\frac{1}{T}}\sum_{\pm}e^{-\sfrac{\pm}{2T}}\ket{\pm}_{H_{1}} \otimes \ket{\pm}_{H_{2}}$, where $H_{1}$ is the Hilbert space of our system and $H_{2}$ is the Hilbert space of the ancilla bits (or the environment) such that $\ket{\Psi} \in H_{1} \otimes H_{2}$ allows purification of $\rho_{0} = Tr_{H_{2}}(\ket{\Psi}\bra{\Psi})$ in the larger Hilbert space. Tracing out $\ket{\Psi}$ over the ancilla qubits leads to $\rho_{0} = \nicefrac{e^{-\beta H_{0}}}{Tr(e^{-\beta H_{0}})}$ where $H_{0} = H_{mix}$ is the mixing Hamiltonian in QAOA; that is, by tracing out we get the thermal state of the mixing Hamiltonian of the QAOA and from which we can reach the thermal state of a target Hamiltonian as mentioned earlier- this is the quantum approximate thermalization. The biggest difference to QAOA is the preparation of the initial state and it is useful to note that the state $\ket{\Psi}$ can be prepared with a low constant depth circuit composed uniquely of $R_{x}$ gates and $CNOT$ gates, hence state preparation is quite efficient \cite{verdon}.

\section{Probabilistic Graphical models (PGMs)}
On a high level, probabilistic graphical models are a graph-based representation of a probability distribution over multiple variables. Rephrasing slightly, PGMs are statistical models that encode complex joint multivariate probability distributions using graphs \cite{hollander}. Each node in the graph corresponds to a random variable, and edges in the graph imply the flow of influence between the variables. In general, specifying a joint probability distribution over nontrivial amount of random variables can be a daunting task. For example for $n$ discrete random variables where each can take $k$ different values, we need a table with $k^{n}$ rows to fully specify the joint probability distribution. As the size grows exponentially with $n$, this approach is clearly infeasible for any reasonably big $n$. However, specifying values of probability for each combination of values of random variables is seldom necessary when dealing with the real-world problems, where it isn’t the case that all the variables are pairwise dependent. If there are independencies between random variables, we can leverage them to express the distribution more concisely. This idea is what is empowering the compact representations provided by probabilistic graphical models.

There are two flavors of PGMs, Directed Graphical Models (DGMs), otherwise known as Bayesian Networks (BNs) and Undirected Graphical Models (UGMs) or Markov Random Fields (MRFs). The main conceptual difference between a Markov network and Bayesian Network is that while casual relationships between the variables in the Bayesian Networks are directed, the relationships between variables in a Markov network are undirected and hence represent a mutual influence of the connected nodes.

By knowing the graph structure of a PGM, one can solve tasks such as inference (computing the marginal distribution of one or more random variables) or learning (estimating the parameters of probability functions). One can even try to learn the structure of the graph itself, given some data. More formal details on inference are covered in Section \ref{morePGM}.

\begin{figure}[h!]
	\centering
	\includegraphics[scale=0.5]{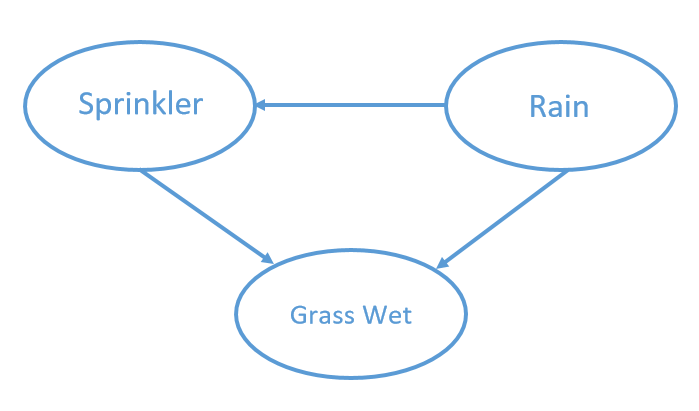}
	\caption{\small{Popular Toy Bayesian Network}}
	\label{fig:toymodel}
\end{figure}

The directed edges (“arrows”) of a BN represent conditional distributions. If the values of the vertices are binary, for example, the conditional distributions may be Bernoulli distributions. In case of continuous values, the conditional distributions may be Gaussian. The joint probability distribution is formulated as a product of conditional or marginal probabilities. One of the popular toy examples of BN is the scenario shown Figure \ref{fig:toymodel}, which encodes the following logic: the probability that the grass is wet is dependent on turning on the sprinkler and the rain. The probability that the sprinkler is on is itself dependent on the rain (you wouldn’t turn on the sprinkler during rain).

This DAG represents the (factorised) probability distribution:
\begin{equation}
p(S,R,G) = p(R)p(S|R)p(G|S,R)
\end{equation}
where $R$ is the random variable for rain, $S$ for the sprinkler and $G$ for the wet grass.

As far as the inference in this scenario is concerned, one can work out the marginal probability $p(G)$ that the grass is wet; in order to calculate this it is required to marginalize the joint probability distribution $p(S,R,G)$
over $S$ and $R$. Another inference task of interest is to calculate the conditional probability that the grass will be wet given that it is not raining. In this case, we would proceed as follows:
\begin{equation}
p(G|R) = \sum_{S}p(S,G|R) = \sum_{S}p(S|R)p(G|S,R)
\end{equation}

For this case we only have to marginalize over $S$ since $R$ is already assumed as given (for example $R=0$  in the case of not raining). This procedure is called variable elimination. Variable elimination is an exact inference algorithm. Its downside is that for large BNs it might be computationally intractable. Approximate inference algorithms such as \textit{Gibbs sampling} or rejection sampling might be used in these cases; more aspects on sampling are covered in Section \ref{QuantumSampling}.

Similar to Bayesian networks, MRFs are used to describe dependencies between random variables using a graph. However, MRFs use undirected instead of directed edges. They may also contain cycles, unlike Bayesian Networks. Thus, MRFs can describe a different set of dependency relationships than their Bayesian network counterparts. Since in this paper, we focus more on MRFs, the relevant information is elaborated in the following section.

\subsection{Markov Networks}
To specify a probability distribution using a Markov network, we need to \textit{parameterize} it. A common parameterization of a Markov network resulting in a joint probability distribution over its variables uses a set of \textit{potential functions}, also referred to as \textit{factors}. Each potential function $E_{k}[C_{k}]$ corresponds to a given clique $C_{k}$ in the graph of the Markov network. It takes in the values of the random variables which form the clique $C_{k}$ and outputs a real number; recollect that we identify random variables with the vertices in the Markov network and this can be performed since there is bijective mapping between the vertices and random variables. To get a probability of a given assignment of variables in a Markov network, we multiply the values of the factor functions and normalize the result. One more side note: a clique of a graph is a complete subgraph of the graph; $0$-cliques correspond to the empty set (sets of $0$ vertices), $1$-cliques correspond to vertices, $2$-cliques to edges, and $3$-cliques to $3$-cycles, etc.

The structure of a Markov network imposes a set of independence assumptions. We refer to these assumptions as conditional, since no two random variables in a connected Markov network are independent unless we condition on an other random variable in the network. $X$ is conditionally independent of $Y$ given $W$ denoted as $(X\bot Y|W)$ if $p(X=x, Y=y | W=w) = p(X=x | W=w)p(Y=y | W=w)$ for all $x \in X, y \in Y, w \in W$. In other words, two random variables conditional independent given a third random variable if we can factorize the joint probability distribution of $X$ and $Y$  given $W$ into individual parts,  probability of $X$ given $W$ and the probability of $Y$ given $W$; It doesn't mean that they are independent, but conditioned on this third random variable, they are independent. The conditional assumptions are generally formalized as Markov properties, and have to be satisfied by every distribution that factorizes over the Markov network. Putting the other way, a fundamental property of MRFs is that they satisfy the pairwise, local and global Markov properties. The \textit{pairwise} Markov property states that two non-neighboring variables are conditionally independent given all other variables:
\begin{equation}
X_{a} \perp X_{b} | X_{G\setminus[a,b]}
\end{equation}
$X_{a}$ and $X_{b}$ defining any two non-neighboring variables and $G_{G}$ being the set of all variables. The \textit{local} Markov property introduces the concept of the \textit{neighborhood} of a variable:
\begin{equation}
X_{a} \perp X_{G \setminus N(a)} | X_{N(a)}
\end{equation}
where $N(a)$ is the neighborhood of $X_{a}$. In other words, any variable is conditionally independent of any other variables given its neighborhood. Finally, the \textit{global Markov property} states that any \textit{set} of variables $X_{A}$ is independent of any other \textit{set} $X_{B}$ given a separating subset $X_{S}$:
\begin{equation}
X_{A} \perp X_{B} | X_{S}
\end{equation}
We say that a set of vertices $S$ \textit{separates} sets $A$ and $B$ in the Markov network, if for every path between the members of $A$ and $B$, all the nodes or vertices of the path does not exist in $S$; if $V(p)$ is the vertices of path $p$, then $V(p) \cap S$ is null set.

In general, the pairwise Markov property follows from the local Markov property, which in turn follows from the global Markov property (global Markov property implies local Markov property). However, for strictly positive probability distributions, the statements are equivalent.

The introduction of cliques in the definition of potential functions in the Markov network can seem arbitrary and counterintuitive. But, it is related to the expressive power of the Markov networks. If we restrict ourselves to much more intuitive concept of edge potentials, in other words, to factors corresponding to the cliques of size 2, it is proven and demonstrated in the literature that it would \textit{not be sufficient} to model a huge class of probability distributions. Of course, it is also to be noted that parameterization using potential functions has certain disadvantages. For one, the structure of the Markov network does not have to correspond to the most granular factorization of the distribution.

Having remarked about the sufficiency of the Markov network model to express particular joint probability distributions, it remains to investigate what class of probability distributions satisfies the Markov network. An important result for MRFs in this regard is captured by the \textit{Hammersley-Clifford theorem}: informally, this theorem states that \textit{a strictly positive probability distribution} that satisfies one (or equivalently all) of the Markov properties may be represented as a Gibbs measure; the Gibbs measure is a strictly positive function factorized over the cliques of the graph:
\begin{equation}
p(X) = \frac{1}{Z} \prod_{k} E_{k}[C_{k}]
\end{equation}
where $Z$ is an appropriate (global) normalization constant (also called the partition function), $C_{k}$s are the cliques of the graph, $E_{k}[C_{k}]$ is the factorized function on the clique (not necessarily normalized) and $X$ is the set of random variables. Needless to say, positive probability distributions cannot model events that have zero probability and the Markov network with this restriction is also referred to as \textit{Gibbs random field} in the literature.

The remark of Gibb’s measure in fact connects to the Gibb’s distribution extensively discussed in \ref{eth}. In other words, a particular probability distribution that is tightly connected with the concept of the Markov network is \textit{Gibbs distribution}, also called as \textit{Boltzmann distribution}. Again, as discussed earlier, the Gibbs distribution has its origins in statistical physics, and it is used to describe the probability of a certain configuration of a physical system (described by the state’s energy) given the temperature of the system, provided that the system is in thermal equillibrium. In this case, the factors $E_{k}[C_{k}]$ are energy defined over the cliques and $p(X) = \frac{1}{Z}\sum_{k} e^{-(E_{k}[C_{k}])}$.

Thus, under mild assumptions, there's a correspondence between Markov networks and the probability distributions they can describe and Boltzmann distributions. It is also to be noted that the Boltzmann distribution is positive. If $p$ is a Boltzmann distribution over a Markov Network $G$, all local Markov properties will hold. The other way also holds if $p$ is a positive distribution.

To summarize, probabilistic graphical models provide very concise way how to specify joint probability distributions over a set of random variables; that is, they are very good at capturing the sparsity structure between random variables. In a graphical model, complexity is dealt with through graph theory. We get both an efficient treatment of uncertainty (probabilities) and of logical structure (independence constraints). Further, they are very explainable and intuitive. 

\subsubsection{Ising Model}\label{ising}
In this section, we briefly consider the special case of MRFs- the Ising model. More details on Ising models can be found in \cite{qudo} and the references therein. The Ising model, named after German physicist Ernst Ising, is a concept that arose in statistical physics and was original inspiration for the concept of a Markov network itself. Ising model describes a abstract physical system of $N$ interacting components (usually magnets, or atoms), arranged according to particular graph structure (most often a $n$-dimensional lattice). Each component can have two states (spins), and \textit{only neighbouring components interact, while strength of the interaction can vary}. Note that in the Ising model, parameterization happens only using only edge potentials ($2$-cliques) and the $1$-clicks; this form of a Markov network is sometimes referred to as \textit{pairwise Markov random field}. Even though it is not as general as using maximum-clique potential functions, Ising model can be used to solve various hard problems, in particular, a Ising formulation for each of 21 Karp’s NP-complete problems exists \cite{babej} with at most cubic overhead in terms of number of spins required respective to the problem size.

More formally, the Ising model assumes that every node can assume two states: $\sigma_{i} = {1, -1}$. The energy of a particular state $\sigma$ of the model (network), which is a string (vector) of states of all the nodes, is described by \cite{hollander}:
\begin{equation}\label{eq:Ising}
E(\sigma) = -\sum_{\braket{i,j}}J_{ij}\sigma_{i}^{z} \sigma_{j}^{z} - \sum_{i}h_{i}\sigma_{i}^{z}
\end{equation}

Note the cliques of a single node (referred to as the on-site fields) and two cliques of two nodes in the model described through the Equation \eqref{eq:Ising}. The joint Gibbs distribution of the system is given by:
\begin{equation}
p(\sigma) = \frac{e^{-E(\sigma)}}{\sum_{\sigma}e^{-E(\sigma)}}
\end{equation}

Taking the logarithm of this expression, the expression factorizes into the cliques of the graph. Thus, we can see that the model is indeed a MRF according to the Hammersley-Clifford theorem. It is also useful to note that the model does not explicitly define the temperature, but it is implicitly there in the constants defining the energy states. One more additional remark- for the Ising model the factors are defined as degree-$1$ and degree-$2$ monomials of the random variables connected in the graph.

Before moving to inference on PGMs, it is to be noted that the quantum mechanical Hamiltonian of the classical Ising model of Equation \eqref{eq:Ising} can be obtained by suitably replacing {$\sigma_{i}$ at the site $i$ by the Pauli operator $\sigma_{i}^{z}$; more details are available again in \cite{qudo} and the references therein.
	
\subsubsection{Remarks on PGM Construction}\label{Remarks}
In the case of Ising Markov networks, the conditional independences are already encapsulated by the model: for instance some spins may not interact (like, spins $0$ and $2$ do not interact, etc). In general, it is hard to learn the structure of a probabilistic graphical model given a set of observed correlations in the data sample $\Omega$. We can only rely on heuristics. The typical way of doing it is to define a scoring function and do some heuristic global optimization.

Once we identified or defined the graph structure $G$, we have to learn the probabilities in the graph. We again rely on our sample and its correlations, and use a maximum likelihood or a maximum a posteriori estimate of the corresponding parameters $\theta_{G}$ with the likelihood $p(\Omega|\theta_{G})$. This is again a hard problem.

\subsection{Additional Points on Inference in PGMs}\label{morePGM}
Once constructed or available, PGMs provide an expressive framework for dealing with complex, large systems of interacting variables. The formalism of probabilistic graphical models is not only useful to capture and concisely express the structure of a given probability distribution, but can be also used to answer questions as introduced in this section. Indeed, it is the aforementioned concise structure of the probabilistic graphical models that allows us to answer questions efficiently, even though they can describe extremely high dimensional probability distributions.

Inference can be loosely described as the problem of deriving a likely state of some unknown random variables in the model based on available information about the state of other, known variables in the model. The known variables in the model are also often referred to as \textit{evidence}. Inference in undirected probabilistic graphical models is often used for problems in machine learning and computer graphics, such as image restoration, image denoising, segmentation or information retrieval.

Three most widely used inference problems are the following: computation of the conditional probability distribution, the most probable explanation (MPE) query and the maximum a-posteriori (MAP) query. We define the queries in the context of Markov networks, but the concepts straightforwardly extend to other probabilistic graphical models or distributions.

Conditional probability distribution query (CPD) arises anytime a subset of the random variables is observed. In the light of this new information, the probability distribution over unobserved variables changes. If $E$ is a subset of variables observed with instantiation $e$, i.e, $E=e$ and the remaining unobserved variables is the subset $Y$, then CPD is about evaluating $p(Y|E=e)$.

The most probable explanation (MPE) problem arises in many complex domains where we have partial information about the state of the system being modelled. Using an MPE query we can figure out what is the most probable state of the system, assuming our partial information is correct. Hence MPE inference is used in situations where filling in the missing information is required, such as image restoration. Again, $E$ if is a set of variables observed with assignment $e$, then the most probable explanation query is defined as the most likely assignment to the set of all unobserved variables $Y$: $\argmax_{\:Y} p(Y|E=e)$.

The maximum-a-posteriori query is a generalization of the most probable explanation query. In the MAP query, one is no longer restricted to ask for the most likely assignment to \textit{all} unobserved variables, but can instead request most likely assignment to a \textit{subset} of the unobserved variables. Formally it is $\argmax_{\:Y} \sum_W p(Y,W|E=e)$, where, $W$ is the set of unobserved variables the query does not include. It is to be noted that if we are only interested in some subset, it is required to marginalize over those variables that we are not interested in; we are only looking for an optimal configuration over a subset that we are interested in. One often given example goes like this - suppose we have a probability distribution describing different symptoms and their correlations and also we have a patient and we observe couple of those symptoms but not others. Then we can run a query to find out all the other symptoms that we shouldn't be looking, assuming that the person has a certain disease. 

In the queries CPD, MPE and MPE, when $E$ is instantiated or assigned with $e$, it is also referred as the the random variables of the set $E$ are \textit{clamped} to the values suggested by $e$.

\subsubsection{More on Complexity of Inference and the Necessity of Sampling}
We mentioned the hardness of PGM model building in \ref{Remarks}. Here, we focus only on the complexity of the inference over PGMs, particularly Markov networks. As noted in the previous sections, the inference involves computing different probability distributions, and generally the exact inferences are are computationally intractable in the worst case. In fact, most of these problems are at least NP hard \cite{wittek}. Considering the special case of Markov networks, the Ising model, the exact inference with them is also usually hard due to the presence of the partition function. As a side note, it is useful to note that computing the partition function lies at heart of many problems in the field of undirected graphical models. In summary, even if we train (build) the network and it reproduces the probability distribution that we are interested in, running the queries mentioned in the previous section is still computationally very difficult.

This suggests the need for approximate inference methods over PGMs. In approximate inference, the task is to approximate the requisite probability distributions, including the approximation of partition function involved. Needless to say, in approximations, we allow some error in the process. One simple formalism can provide a feel: for example, given a value of a random variable $X$ can take, finding a value $\tau$ such that $|P(X=x)-\tau | \leq \epsilon$, where, $\epsilon$ is the allowable error can be seen as approximate inference.

One popular approach to carry out approximate inference in PGMs is through the usage of sampling-based methods. The main idea of \textit{sampling-based} methods is to avoid the need for computing the joint probability distribution by approximating it using a set of realizations of the variables in the network. It is useful to note that sampling methods are also capable of handling the state space explosion in high-dimensional probabilistic graphical models. In short, in carrying out the inference over PGMs, instead of solving it accurately, we can run some sampling over possible outcomes and use them for approximate inference. This raises the issue of how to sample efficiently. One popular approach is Markov Chain Monte Carlo (MCM) sampling, which is briefed next.

\subsubsection{Markov Chain Monte Carlo Sampling}
As its name suggests, Markov chain Monte Carlo sampling techniques possess two main properties. First, they use a Markov Chain to generate random samples using a particular sequential process. Monte Carlo part of the name is a reference to a particular family of randomized algoritms, unsuprisingly called Monte Carlo algorithms. Algorithms that are members of this family can produce output that is incorrect with some, typically small, probability (unlike Las Vegas randomized algorithms, where the correctness of the output is guaranteed). Hence, the second property is the fact that Markov Chain Monte Carlo is a method of not an exact, but approximate sampling.

Using Markov Chain Monte Carlo methods we guarantee the ability to sample from the target distribution directly. We use a stochastic process that samples from probability distributions that are gradually approaching the target distribution. A Markov chain is a time indexed random variable, where at each time instant the random variable represents state of all of the variables of the network. Further, they have \textit{memoryless} property, namely, the probability of achieving the next state depends only on the current state, and not on any previous state of the process. Many a times, we restrict ourselves to homogeneous Markov chains, which ensures that the dynamics of the system do not change with time. The premise of using Markov chains is in the fact that they can, under certain conditions, eventually reach an equilibrium state, where the probability distribution over the next state of the Markov chain is equal to the probability distribution over the current state, referred to as equilibrium distribution. Of course, convergence to stationary distribution, or its uniqueness are not guaranteed in the general case. In summary, MCMC methods attempt to simulate direct draws from some complex distribution of interest and they use the previous sample values to randomly generate the next sample value, generating a Markov chain (as the transition probabilities between sample values are only a function of the most recent sample value).

In practice, the MCMC approach can suffer from speed, accuracy, or both. There are two main issues that pose roadblocks to classical MCMC sampling algorithms. First main issue, so called \textit{burn in} phase lies with the fact that MCMC sampling initially does not sample from the target distribution directly- initial states are sampled from an arbitrary initial distribution, which only slowly converges to target. The amount of time the Markov chain needs to converge to the target (stationary) distribution is called the \textit{mixing time} or \textit{burn in} time. Second issue is the \textit{autocorrelation} of the subsequent samples from the Markov chain Monte Carlo process, which inherently stems from the production of samples via transitions in the Markov chain over the state space of all possible instantiations of the random variables.

\section{Quantum Enhanced Sampling for Probabilistic Graphical Inference}\label{QuantumSampling}
The roots of probabilistic graphical models go back to the 1980s, with a strong connection to Bayesian statistics. The story resembles that of neural networks: they have been around for over three decades and they need massive computational power. However, unlike in the case of deep learning, the requirements for computational resources remain out of reach. These models require sampling a distribution, and very often it is the Boltzmann distribution. Given that Markov Chain Monte Carlo sampling techniques have performance issues (slow burn-in time and correlated samples), the development of new methods for sampling high-dimensional Boltzmann distributions is paramount to adoption of probabilistic graphical models in areas such as machine learning, where performance is critical. Since quantum computers can give samples from this distribution, we can hope that quantum hardware can enable these models the same way graphics processing units enabled deep learning. The potential of using quantum resources in machine learning problems that go beyond simple speedups and that give an advantage in the quality of the models. Large data sets are pushing computational abilities of current and future classical systems to their limit. Much as the GPU accelerated computing in the past decade, the research groups and industry alike are looking at quantum machine learning as the new paradigm to enable previously intractable computational capabilities. The viability of using current quantum computing architectures for the purposes of sampling, which is an inherent component of inference and training of many machine learning models, is already getting demonstrated. The existing results suggest that while current quantum annealing architectures are capable of achieving comparable sampling quality as classical algorithms, they also outperform simple classical algorithms on conventional desktop stations significantly. Substantial robustness of the system against noise, which implies engineering maturation of the device is also reported. The main promise of quantum-enhanced sampling lies in the enabling models such as large-scale probabilistic models, or deep Boltzmann machines. While these feats are yet unrealized, the current state of the field and speed of innovation within it hints at possible bright future ahead.

The discussions so far bring one more point in contrast to deep learning: once the neural network is trained, running a prediction on it is relatively cheap. In the case of probabilistic graphical models, inference remains computationally demanding even after training the model. As already mentioned instead of solving the inference problem directly, we use approximate inference with sampling and this is exactly the step we can replace by sampling on a quantum computer. Also, as captured previously, the Markov networks through their clicks factorizes a Boltzmann distribution. In the special case of binary random variables, we can set up an Ising model and use quantum computers to do approximate inference. Presently, we have options. If we want to use quantum annealer, we have a fair number of qubits and our task is to estimate the effective temperature of the system, so you can rescale these energy values according to what the hardware is actually executing; more details are already covered in Section \ref{Gibbs}. Or we can run the gate model quantum approximate thermalization protocol, which needs an ancilla system, which means that it is required to use a lot more qubits than what we are actually interested in. On a smaller scale, this is an option too; more details in Section \ref{QAT}. Whichever method we choose, we can accelerate some of these Markov networks at training and also during inference.

Before concluding the descriptive part of this paper, it is useful to note that graphical models are quintessentially generative: we explicitly model a probability distribution. Thus generating new samples is trivial and we can always introduce extra random variables to ensure certain properties. These models also take us a step closer to explainability (as already mentioned), either by the use of the random variables directly for explanations (if your model is such) or by introducing explanatory random variables that correlate with the others.

We have already considered Ising models, which are the special Markov networks. In our onging work, we are looking into Boltzmann machines, which are in turn special class of Ising models. Here, we partition the Ising spins into two categories- visible nodes and others are called hidden nodes. For Boltzmann machines, the energy function is similar to ordinary Markov network, but we are only interested in reproducing some probability distribution here on the visible nodes. The hidden nodes are there to help mitigate correlations between the visible random variables. So we marginalize out over the hidden nodes to get a probability distribution that we are interested in. These are very powerful methods, and they are expensive to train on classical computers. So this is one area where quantum computers can help.

\section{Results and Discussion}
Few typical results and remarks from our elaborate simulation studies with different networks and clamping scenarios are recorded in this section. As discussed previously, two popular quantum computing paradigms, Quantum Annealing and Quantum Approximate Optimization Algorithm, have been used to draw out samples from the simple $3$ and $4$ node Markov networks shown in Figures \ref{fig:3graph} and \ref{fig:4graph}, repectively. From the samples obtained it was possible to construct the joint probability distribution of the variables represented by the nodes of the graph. Ideally, the frequency of obtaining each sample should match the probability predicted by the Boltzmann distribution:
\begin{equation}\label{eq:Boltzmann}
p(E_{i}) = \nicefrac{g(E_{i}) e^{-\nicefrac{E_{i}}{T}}}{Z}
\end{equation}
where $E_{i}$ are the energy of the configurations, $T$ is the temperature of the bath, $g(E_{i})$ is the degeneracy of the energy $E_{i}$, and $Z$ is, as usual, the partition function.

The distribution of samples and the Boltzmann distribution \ref{eq:Boltzmann} have been compared through Kullback-Liebler (KL) divergence. Towards having proper comparison, more samples were considered by using a large number of ``shots" in the experimentation. Further, we could marginalize over some of the variables and find the distribution of the remaining, find the Maximum A Posteriori (MAP) estimation for one or more variables, and also clamp over some of the variables and check the effect of such clamping over the other variables.
\begin{figure}
	\centering
	\begin{subfigure}[b]{0.49\textwidth}
		\centering
		\includegraphics[scale=0.5]{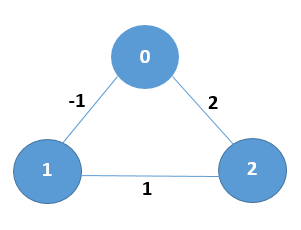}
		\caption{}  
		\label{fig:3graph}
	\end{subfigure}
	\hfill
	\begin{subfigure}[b]{0.49\textwidth}  
		\centering 
		\includegraphics[scale=0.5]{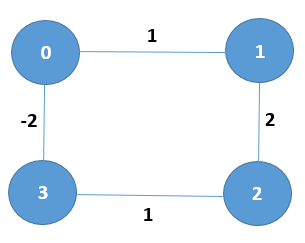}
		\caption{}
		\label{fig:4graph}
	\end{subfigure}
	\caption{\small{Simple Markov networks for sampling and inferencing.}}
	\label{fig:graphs}
\end{figure}

Figures \ref{fig:n3T3} and \ref{fig:n3T1000} show the frequency of the states sampled from the $3$ node graph in Figure \ref{fig:3graph} using the simulated annealing API provided by dimod \cite{dimod} at low and high temperatures respectively, against their energy, along with the Boltzmann distribution for comaprison. The similarity of the two distributions has been portrayed in Figure \ref{fig:kld} through their KL divergence metric for different values of temperature. It must be noted that simulated samples were used to compensate for the lack of access to a Quantum Annealing hardware. If an annealer, like the D-Wave machine is available, the requisite samples can be drawn from it.
\begin{figure}
	\centering
	\begin{subfigure}[b]{0.49\textwidth}
		\centering
		\includegraphics[scale=0.5]{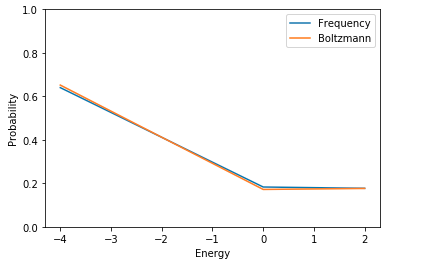}
		\caption{}  
		\label{fig:n3T3}
	\end{subfigure}
	\hfill
	\begin{subfigure}[b]{0.49\textwidth}
		\centering
		\includegraphics[scale=0.5]{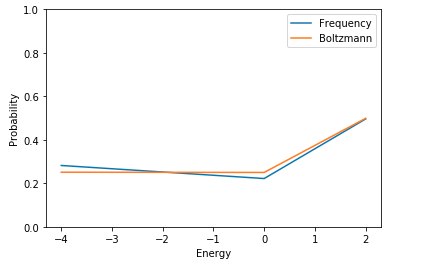}
		\caption{}  
		\label{fig:n3T1000}
	\end{subfigure}
	\caption{\small{Subfigure (a) compares the obtained frequency and Boltzmann distributions for different configurations corresponding to their respective energy. (b) Shows the same at a much higher temperature.}}
\end{figure}

\begin{figure}
	\centering 
	\includegraphics[scale=0.5]{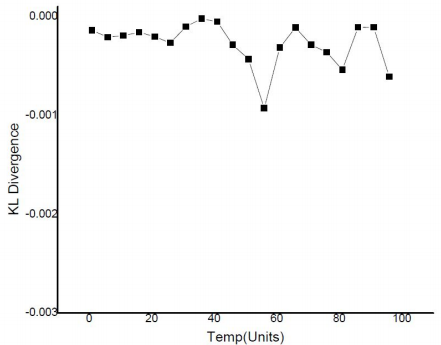}
	\caption{\small{Figure shows the variation of KL Divergence between the frequency and Boltzmann distributions at different temperatures.}}
	\label{fig:kld}
\end{figure}

As evidenced by Figure \ref{fig:kld}, the KL Divergence of the frequency and Boltzmann/Gibbs distributions are always close to zero, signifying the fact that samples obtained from Quantum Annealing do indeed follow the requisite distribution at all temperatures. The joint probability distributions of the $3$ node Markov network in Figure \ref{fig:3graph} at temperatures $T=3$ and $T=1000$ have been detailed in Table \ref{table:n3distribution}.

\begin{table}[h!]
	\centering
	\begin{tabular}{|c|c|c|} 
		\hline
		Configuration & Probability at $T=3$ & Probability at $T=1000$ \\ [0.5ex] 
		\hline
		-1,-1,-1 & 0.04410419473467003 & 0.12474987533391639\\
		-1,-1,1 & 0.3258883690926387 & 0.125500624581416\\
		-1,1,-1 & 0.04410419473467003 & 0.12474987533391639\\
		-1,1,1 & 0.08590324143802122 & 0.12499962475075126\\
		1,-1,-1 & 0.08590324143802122 & 0.12499962475075126\\
		1,-1,1 & 0.04410419473467003 & 0.12474987533391639\\
		1,1,-1 & 0.3258883690926387 & 0.125500624581416\\
		1,1,1 & 0.04410419473467003 & 0.12474987533391639\\
		\hline
	\end{tabular}
	\caption{\small{Joint Probability Distribution for graph in Figure \ref{fig:3graph}}}
	\label{table:n3distribution}
\end{table}

Having established the authenticity of the samples drawn, we proceed to draw inference from the obtained probability distributions by marginalizing over the last variable and finding the Maximum A Posteriori estimate for the last variable, the results for both of which for the $3$-node graph can be found in Tables \ref{table:n3T3MarMAP} and \ref{table:n3T1000MarMAP}, for temperatures $T=3$ and $T=1000$, respectively. Further, the marginalized distributions have been captured in Figures \ref{fig:n3T3Mar} and \ref{fig:n3T1000Mar}. As expected, the distribution in Figure \ref{fig:n3T3Mar} is flat, such that all the states have equal probability of occurring, at high temperature ranges. 

\begin{table}[h!]
	\centering
	\begin{tabular}{|c|c|c|} 
		\hline
		Configuration & Marginalized Distribution & MAP Estimate \\ [0.5ex] 
		\hline
		-1,-1 & 0.3699925638273087 & 1\\
		-1,1 & 0.13000743617269125 & 1\\
		1,-1 & 0.13000743617269125 & -1\\
		1,1 & 0.3699925638273087 & -1\\
		\hline
	\end{tabular}
	\caption{\small{Results of marginalization and Maximum A Posteriori estimate over the last variable in Figure \ref{fig:3graph} at $T=3$}}
	\label{table:n3T3MarMAP}
\end{table}

\begin{table}[h!]
	\centering
	\begin{tabular}{|c|c|c|} 
		\hline
		Configuration & Marginalized Distribution & MAP Estimate \\ [0.5ex] 
		\hline
		-1,-1 & 0.2501504999153324 & 1\\
		-1,1 & 0.24974950008466765 & 1\\
		1,-1 & 0.24974950008466765 & -1\\
		1,1 & 0.2501504999153324 & -1\\
		\hline
	\end{tabular}
	\caption{\small{Results of marginalization and Maximum A Posteriori estimate over the last variable in Figure \ref{fig:3graph} at $T=1000$}}
	\label{table:n3T1000MarMAP}
\end{table}

\begin{figure}
	\centering
	\begin{subfigure}[b]{0.49\textwidth}
		\centering
		\includegraphics[scale=0.5]{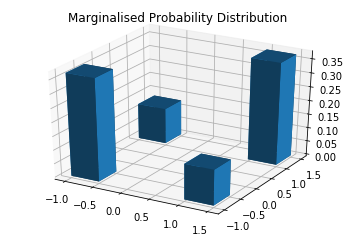}
		\caption{}  
		\label{fig:n3T3Mar}
	\end{subfigure}
	\hfill
	\begin{subfigure}[b]{0.49\textwidth}
		\centering
		\includegraphics[scale=0.5]{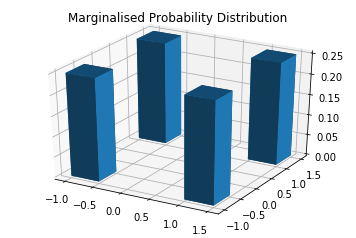}
		\caption{}  
		\label{fig:n3T1000Mar}
	\end{subfigure}
	\caption{\small{Figures (a) and (b) show the distribution for the $3$-node graph in Figure \ref{fig:3graph} after marginalizing over the last variable at temperatures $T=3$ and $T=1000$, respectively}}
\end{figure}

The procedure described above was repeated for the $4$-node graph. The comparison of the frequency distribution with the Gibbs distribution at temperatures $T=3$ and $T=1000$ can be found in Figures \ref{fig:n4T3} and \ref{fig:n4T1000}. The corresponding joint probaility distribution, distribution on marginalising over the last variable and MAP of the last variable at the two temperature values have been shown in Tables \ref{table:n4distribution}, \ref{table:n4T3MarMAP} and \ref{table:n4T1000MarMAP}.

\begin{figure}[h]
	\centering
	\begin{subfigure}[b]{0.49\textwidth}
		\centering
		\includegraphics[scale=0.5]{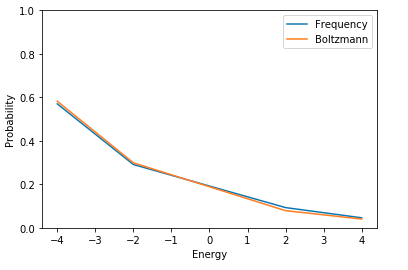}
		\caption{}  
		\label{fig:n4T3}
	\end{subfigure}
	\hfill
	\begin{subfigure}[b]{0.49\textwidth}  
		\centering 
		\includegraphics[scale=0.5]{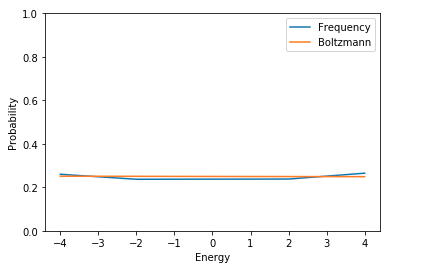}
		\caption{}
		\label{fig:n4T1000}
	\end{subfigure}
	\caption{\small{Sugfigures (a) and (b) show the probability distribution with respect to enrgy for the $4$-node graph at temperatures $T=3$ and $T=1000$, respectively.}}
	\label{fig:n4}
\end{figure}

\begin{table}[h!]
	\centering
	\begin{tabular}{|c|c|c|} 
		\hline
		Configuration & Probability at $T=3$ & Probability at $T=1000$ \\ [0.5ex] 
		\hline
		-1,-1,-1,-1 & 0.0747011997784186 & 0.06262481195896036\\
		-1,-1,-1,1 & 0.14549806971605195 & 0.06275018691604371\\
		-1,-1,1,-1 & 0.010109708030127073 & 0.062250188082706305\\
		-1,-1,1,1 & 0.0747011997784186 & 0.06262481195896036\\
		-1,1,-1,-1 & 0.010109708030127073 & 0.062250188082706305\\
		-1,1,-1,1 & 0.01969192247540231 & 0.06237481304228965\\
		-1,1,1,-1 & 0.01969192247540231 & 0.06237481304228965\\
		-1,1,1,1 & 0.14549806971605195 & 0.06275018691604371\\
		1,-1,-1,-1 & 0.14549806971605195 & 0.06275018691604371\\
		1,-1,-1,1 & 0.01969192247540231 & 0.06237481304228965\\
		1,-1,1,-1 & 0.01969192247540231 & 0.06237481304228965\\
		1,-1,1,1 & 0.010109708030127073 & 0.062250188082706305\\
		1,1,-1,-1 & 0.0747011997784186 & 0.06262481195896036\\
		1,1,-1,1 & 0.010109708030127073 & 0.062250188082706305\\
		1,1,1,-1 & 0.14549806971605195 & 0.06275018691604371\\
		1,1,1,1 & 0.0747011997784186 & 0.06262481195896036\\
		\hline
	\end{tabular}
	\caption{\small{Joint Probability Distribution for graph in Figure \ref{fig:4graph}}}
	\label{table:n4distribution}
\end{table}

\begin{table}[h!]
	\centering
	\begin{tabular}{|c|c|c|} 
		\hline
		Configuration & Marginalized Distribution & MAP Estimate \\ [0.5ex] 
		\hline
		-1,-1,-1 & 0.22019926949447055 & -1\\
		-1,-1,1 & 0.08481090780854568 & 1\\
		-1,1,-1 & 0.029800730505529383 & -1\\
		-1,1,1 & 0.16518909219145425 & 1\\
		1,-1,-1 & 0.16518909219145425 & -1\\
		1,-1,1 & 0.029800730505529383 & 1\\
		1,1,-1 & 0.08481090780854568 & -1\\
		1,1,1 & 0.22019926949447055 & 1\\
		\hline
	\end{tabular}
	\caption{\small{Results of marginalization and Maximum A Posteriori estimate over the last variable in Figure \ref{fig:4graph} at $T=3$}}
	\label{table:n4T3MarMAP}
\end{table}

\begin{table}[h!]
	\centering
	\begin{tabular}{|c|c|c|} 
		\hline
		Configuration & Marginalized Distribution & MAP Estimate \\ [0.5ex] 
		\hline
		-1,-1,-1 & 0.12512499995833337 & -1\\
		-1,-1,1 & 0.12487500004166666 & 1\\
		-1,1,-1 & 0.12487500004166666 & -1\\
		-1,1,1 & 0.12512499995833337 & 1\\
		1,-1,-1 & 0.12512499995833337 & -1\\
		1,-1,1 & 0.12487500004166666 & 1\\
		1,1,-1 & 0.12487500004166666 & -1\\
		1,1,1 & 0.12512499995833337 & 1\\
		\hline
	\end{tabular}
	\caption{\small{Results of marginalization and Maximum A Posteriori estimate over the last variable in Figure \ref{fig:4graph} at $T=1000$}}
	\label{table:n4T1000MarMAP}
\end{table}

Additionally, the conditional effect of one of the variables over the others was observed by clamping the former to a spin state, $+1$ or $-1$ and observing the resultant spin states of the other variables, as elaborated towards the end of Section \ref{morePGM}. To illustrate this, the thermalization procedure was carried out after fixing a vertex of the $4$-node to the $-1$ state and ten thousand samples were drawn from it. From the samples thus obtained, the frequency of the states of the other vertices was observed to draw some conclusion about the influence of the clamped node had on the others. The experimental results can be found in Figures \ref{fig:clamping1} and \ref{fig:clamping2}.

\begin{figure}
	\centering
	\begin{subfigure}[b]{0.49\textwidth}
		\centering
		\includegraphics[scale=0.358]{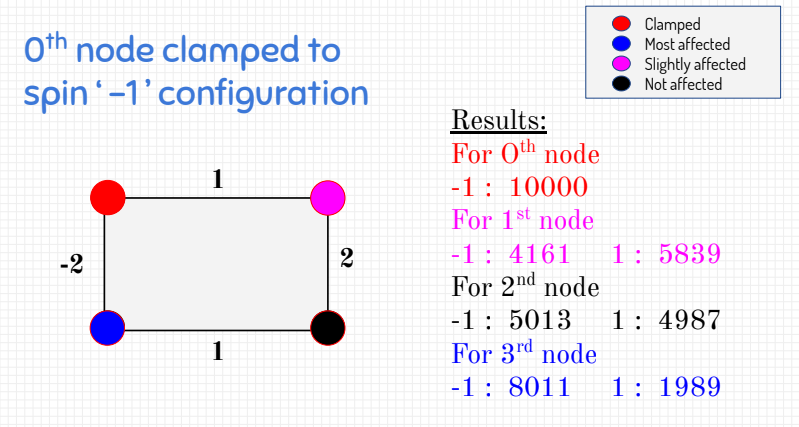}
		\caption{}  
		\label{fig:clamping1}
	\end{subfigure}
	\hfill
	\begin{subfigure}[b]{0.49\textwidth}
		\centering
		\includegraphics[scale=0.5]{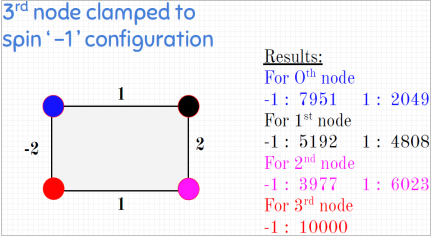}
		\caption{}  
		\label{fig:clamping2}
	\end{subfigure}
	\caption{\small{Subfigure (a) shows the effect of fixing the $0^{th}$ node of the $4$-node graph to the state $-1$ on the states of the other vertices. In Subfigure (b) the process was repeated on the $3^{rd}$ node.}}
\end{figure}

\begin{figure}
	\centering
	\begin{subfigure}[b]{0.49\textwidth}
		\centering
		\includegraphics[scale=0.5]{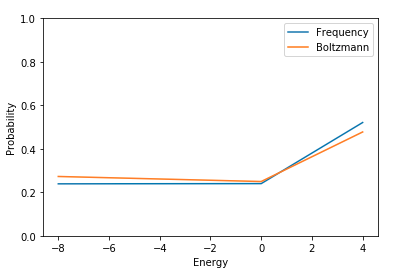}
		\caption{}  
		\label{fig:n3T90}
	\end{subfigure}
	\hfill
	\begin{subfigure}[b]{0.49\textwidth}
		\centering
		\includegraphics[scale=0.5]{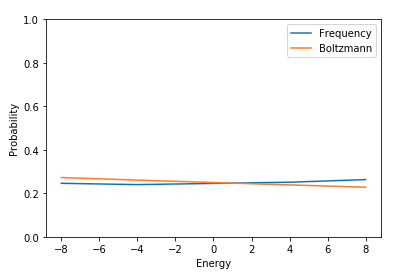}
		\caption{}  
		\label{fig:n4T90}
	\end{subfigure}
	\caption{\small{Subfigure (a) compares the obtained frequency (using Quantum Approximate Thermalization) and Boltzmann distributions for different configurations of the $3$-node graph corresponding to their respective energy at $T=90$. (b) Shows the same for the $4$-node graph.}}
\end{figure}

Samples were also drawn using Quantum Approximate Thermalization, described in Section \ref{QAT}. In doing so, the QAOA routine from Qiskit was used and the thermal state of the cost Hamiltonian had been approached through the usual unitary pair applications  \cite{code,wittekcode,qiskit}. As mentioned in Section \ref{QAT}, the circuitry of the purified initial state, involving the qubits of the subsystem and the ancilla was utilized in the routine. The unitary evolution happens over the subsystem qubits, leading to the required thermal states, starting from the thermal state of the mixing Hamiltonian. The samples obtained for a high temperature have been portrayed in Figures \ref{fig:n3T90} and \ref{fig:n4T90}, for the $3$ and $4$-node graphs, respectively. The results obtained using this method at low temperatures differed from the corresponding Gibbs distribution and warrants further investigation. The simulations were carried out using \cite{wittek,code,wittekcode}. To illustrate the results, the distribution of samples for the $3$-node graph at $T=0.5$ and $T=5$ have been included in Figures \ref{fig:n3T_5} and \ref{fig:n3T5}, repectively. In the experiment carried out to obtain these figures, the number of shots was set at one thousand, and the number of times configurations with given energy values were obtained have been plot as is. One can obtain the probability of obtaining each energy state by dividing this number by the total number of shots.

\begin{figure}
	\centering
	\begin{subfigure}[b]{0.49\textwidth}
		\centering
		\includegraphics[scale=0.5]{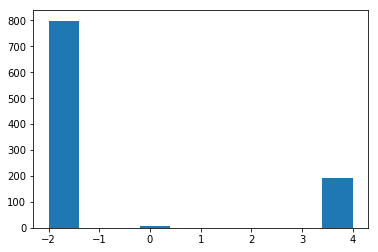}
		\caption{}  
		\label{fig:n3T_5}
	\end{subfigure}
	\hfill
	\begin{subfigure}[b]{0.49\textwidth}
		\centering
		\includegraphics[scale=0.5]{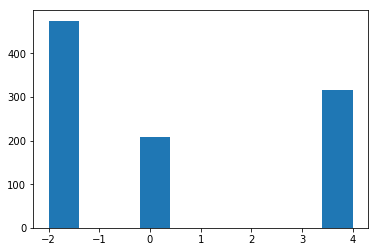}
		\caption{}  
		\label{fig:n3T5}
	\end{subfigure}
	\caption{\small{The subfigures (a) and (b) show the number of times states corresponding to each energy value were obtained on the y-axis, and the energy values on the x-axis, for the $3$-node graph at $T=0.5$ and $T=5$, respectively.}}
\end{figure}

\section{Conclusion}
In this paper, an attempt was made to provide useful notes linking quantum thermalization, sampling from the quantum setups and using them for inference on probabilistic graphical models. Simulation results were suitably included to further elaborating the concepts in this useful topic relevant to Quantum Machine Learning. Work is in progress towards a systematic study of Restricted Boltzmann Machines by incorporating appropriate quantum algorithmic primitives and approaches.

\end{document}